\documentclass[11pt]{article}
\usepackage[english]{babel}
\usepackage{amsmath} \usepackage{amssymb} \usepackage{bm}
\RequirePackage[numbers,sort&compress]{natbib}
\usepackage{color}

\def\be{\begin{equation}}
\def\ee{\end{equation}}
\def\onehalf{{\textstyle{\frac{1}{2}}}}

\begin{document}

\begin{center}
{\Large \bf Entropy in locally-de Sitter spacetimes}
\vskip 0.5cm
{\bf A. Araujo and J. G. Pereira}
\vskip 0.1cm
{\it Instituto de F\'{\i}sica Te\'orica, Universidade Estadual Paulista \\
Rua Dr. Bento Teobaldo Ferraz 271 \\
01156-970 S\~ao Paulo, Brazil}
\end{center}

\vskip 0.3cm
\begin{quote}
{\footnotesize As quotient spaces, Minkowski and de Sitter are fundamental spacetimes in the sense that they are known {\it a priori}, independently of Einstein equation. They represent different non-gravitational backgrounds for the construction of physical theories. If general relativity is constructed on a de Sitter spacetime, the underlying kinematics will no longer be ruled by Poincar\'e, but by the de Sitter group. In this case the definition of diffeomorphism changes, producing concomitant changes in the notions of energy and entropy. These changes are explicitly discussed for the case of the Schwarzschild solution, in which the black hole and the de Sitter horizons show up as a unique entangled system. Such entanglement, together with energy conservation, create a constraint between the black hole activity and the evolution of the de Sitter radius, providing a new scenario for the study of cosmology.

}
\end{quote}
\vskip 1cm

%
%
%
%
%
%
%
%
%
%
%

\section{Introduction}
\label{intro}

Minkowski and de Sitter spacetimes represent different non-gravitational backgrounds for the construction of physical theories. General relativity, for instance, can be construct on any one of them. Of course, in either case gravitation will have the same dynamics, only their local kinematics will be different. If the underlying spacetime is Minkowski, the local kinematics will be ruled by the Poincar\'e group of ordinary special relativity. If the underlying spacetime is de Sitter, the local kinematics will be ruled by the de Sitter group, which amounts then to replace ordinary special relativity by a de Sitter-ruled special relativity \cite{dSsr0,dSsr1,dSsr2}. In this case, spacetime will no longer present a locally-Minkowski Riemannian structure, but will be described by a Cartan geometry \cite{CartanGeo} that reduces locally to de Sitter---usually called de Sitter-Cartan geometry \cite{wise,hendrik}. Accordingly, in a locally inertial frame, where inertial effects exactly compensate for gravitation, the spacetime metric will reduce to the de Sitter metric.

As is well-known, Minkowski is transitive under spacetime translations. In a locally-Minkowski spacetime, therefore, diffeomorphisms will be inherently related to translations. This is the reason why the invariance of any source field Lagrangian under diffeomorphisms is connected, by Noether's theorem, to energy-momentum conservation. On the other hand, the de Sitter spacetime is transitive under a combination of translations and proper conformal transformations \cite{dSsrJGP}. In a locally-de Sitter spacetime, therefore, the notion of diffeomorphism changes in the sense that it will be inherently related to a combination of translations and proper conformal transformations. The invariance of a physical system under such transformation, according to Noether's theorem, gives rise to a conserved current which, in addition to the usual translational-related energy-momentum current, includes also a piece related to the proper conformal transformation---the so-called proper conformal current \cite{ColemanErice}.

Now, the entropy of a general Killing horizon can be expressed in terms of the Noether charge related to the invariance of a physical system under spacetime diffeomorphisms \cite{WaldEntro,pady2}. As a consequence, in a locally-de Sitter spacetime the very notion of entropy will also change: in addition to the usual entropy related to the translational notion of energy, it will include also a piece related to the proper conformal current---a kind of {\em proper conformal entropy}. The purpose of this paper is to show how this new notion of entropy emerges from first principles, and to explore possible consequences for the physics of black holes and for cosmology.

\section{Basics of de Sitter spacetime}

The de Sitter spacetime can be seen as a hypersurface in the host pseudo-Euclidean space with metric $\eta_{AB}$ = $(+1,-1,-1,-1,-1)$ ($A, B, ... = 0, \dots, 4)$, whose points in Cartesian coordinates $\chi^A$ satisfy \cite{HE}
\be
\eta_{AB} \, \chi^A \chi^B = - \, l^2
\label{dspace2}
\ee
with $l$ the de Sitter length-parameter. The four-dimensional stereographic coordinates $\{x^\mu\}$ are obtained through a ste\-re\-o\-graphic projection from the de Sitter hypersurface into a target Minkowski spacetime. They are defined by \cite{gursey}
\be
\chi^{\mu} = \Omega(x) \, x^\mu
\quad \mbox{and} \quad
\chi^4 = - \, l \, \Omega(x) \left(1 +
{\sigma^2}/{4 l^2} \right)
\ee
where
\be
\Omega(x) = (1 - {\sigma^2}/{4 l^2})^{-1}
\label{n}
\ee
with $\sigma^2$ the Lorentz invariant quadratic form $\sigma^2 = \eta_{\mu \nu} \, x^\mu x^\nu$. In Cartesian coordinates $\chi^A$, the generators of the infinitesimal de Sitter transformations are written as
\be
L_{A B} = \eta_{AC} \, \chi^C \, \frac{\partial}{\partial \chi^B} -
\eta_{BC} \, \chi^C \, \frac{\partial}{\partial \chi^A}.
\label{dsgene}
\ee
In terms of the stereographic coordinates $\{x^\mu\}$, these generators decompose in the Lorentz generators
\be
L_{\mu \nu} =
\eta_{\mu \rho} \, x^\rho \, P_\nu - \eta_{\nu \rho} \, x^\rho \,
P_\mu
\label{dslore}
\ee
and in the de Sitter ``translation'' generators
\be
\Pi_\mu \equiv \frac{L_{4 \mu}}{l} = P_\mu - \frac{1}{4 l^2} \, K_\mu
\label{dStra}
\ee
where
\be
P_\mu = \partial_\mu \quad {\rm and} \quad
K_\mu = \left(2 \eta_{\mu \nu} x^\nu x^\rho - \sigma^2
	\delta_{\mu}^{\rho}
\right) \partial_\rho
\label{cp2}
\ee
are, respectively, the generators of translations and {proper} conformal transformations \cite{coleman}. Although the isotropy of both Minkowski and de Sitter is described by the Lorentz group, their homogeneity properties are completely different: whereas Minkowski is transitive under spacetime translations, we see from Eq.~(\ref{dStra}) that the de Sitter spacetime is transitive under a combination of translations and proper conformal transformations---usually called de Sitter ``translations''.

The de Sitter generators (\ref{dslore}) and (\ref{dStra}) can be rewritten in the form
\be
L_{\mu \nu} = \xi^{~\gamma}_{(\mu \nu)} \, \partial_\gamma
\label{dsloreBis}
\ee
and
\be
\Pi_\mu = \xi^{\;\gamma}_{(\mu)} \, \partial_\gamma
\label{dStraBis}
\ee
where
\be
\xi^{~\gamma}_{(\alpha \beta)} =
\eta_{\alpha \delta} x^\delta \delta^\gamma_\beta -
\eta_{\beta \delta} x^\delta \delta^\gamma_\alpha
\label{LoreKiVec}
\ee
are the Killing vectors of the Lorentz group, and
\be
\xi^{\;\gamma}_{(\alpha)} =
\delta^\gamma_\alpha -
\frac{1}{4l^2}(2 \eta_{\alpha \delta} x^\delta  x^\gamma -
\sigma^2 \delta^\gamma_\alpha)
\label{pia}
\ee
are the de Sitter ``translation'' Killing vectors. In terms of these vectors, a Lorentz transformation assumes the form
\be
\delta_{L} x^\gamma = \onehalf \, \xi^{~\gamma}_{(\alpha \beta)} \,
\varepsilon^{\alpha \beta}(x)
\label{LoreTrans}
\ee
with $\varepsilon^{\alpha \beta}(x) = - \varepsilon^{\beta\alpha}(x)$ the transformation parameters. The de Sitter ``translation'', on the other hand, is written as
\be
\delta_\Pi x^\gamma = \xi^{\;\gamma}_{(\alpha)} \, \varepsilon^\alpha(x)
\label{dStransCoord5}
\ee
with $\varepsilon^\alpha(x)$ the transformation parameters. In the contraction limit $l \to \infty$ it reduces to the ordinary Poincar\'e translations
\be
\delta_{\rm P} x^\gamma = \delta^{\;\gamma}_{(\alpha)} \, \varepsilon^\alpha(x)
\label{Ptrans}
\ee
with $\delta^{\;\gamma}_{(\alpha)}$ standing for the corresponding Killing vectors.

\section{Conservation laws}
\label{Conser}

Let us consider a Cartan spacetime that reduces locally to de Sitter \cite{wise,hendrik}. Invariance of a physical system under the de Sitter ``translations" (\ref{dStransCoord5}) yields, through Noether's theorem, the conservation law \cite{IJMP14}
\be
\nabla_\mu \Pi^{\rho \mu} = 0
\label{UniCon}
\ee
where the conserved current, when written in stereographic coordinates, has the form
\be
\Pi^{\rho \mu} \equiv \xi^{\rho}_{\alpha} \, T^{\alpha \mu} =
T^{\rho \mu} - (2l)^{-2} \, {K}^{\rho \mu}
\label{TmK}
\ee
with $T^{\rho \mu}$ the ordinary symmetric energy-momentum current and
\be
K^{\rho \mu} =
\left(2 \eta_{\alpha \gamma} \, x^\gamma x^\rho -
\sigma^2 \delta_\alpha^\rho \right) T^{\alpha \mu}
\label{KdelT}
\ee
the proper conformal current \cite{ColemanErice}. The current $\Pi^{\rho \mu}$ represents a generalised notion of energy-momentum, which is consistent with the local isotropy of spacetime \cite{almeira}. In particular, the generalised notion of energy will be
\be
E \equiv \Pi^{00} = E_T - (2l)^{-2} \, E_K
\label{geneEner}
\ee
where $E_T$ is the ordinary, translational notion of energy and $E_K$ is the proper conformal notion of energy. Note that now ordinary energy-momentum is allowed to transform into proper conformal current and vice versa, while keeping the total current conserved. This is an important additional freedom, which is not present in locally-Minkowski spacetimes. In fact, in the formal limit $l \to \infty$, the underlying de Sitter spacetime contracts to Minkowski, and we recover the ordinary conservation law of locally-Minkowski spacetimes
\be
\nabla_\mu T^{\rho \mu} = 0.
\label{Conser2}
\ee
Note that the decomposition of $\Pi^{\rho \mu}$ into energy-momentum and proper conformal currents is manifest only in stereographic coordinates. In any other coordinate system, the decomposition (\ref{TmK}) will have a different form.

\section{Entropy in de Sitter-Cartan geometry}
\label{WaldEnSec}

The event horizon of a stationary black hole is a Killing horizon ${\mathcal H}$, that is, a null surface to which a Killing vector field $\zeta^\alpha$ is normal. The surface gravity $\kappa$ at any point of the Killing horizon, defined by the condition
\be
\zeta^\alpha \nabla_\alpha \zeta^\beta = \kappa \, \zeta^\beta,
\ee
determines the physical temperature $T = \kappa/2 \pi$ at the horizon. In the case of spacetimes that reduce locally to Minkowski, which is transitive under translations, the Killing vectors corresponding to this transformation are
\be
\zeta^\gamma \equiv \delta^\gamma = \delta^{\gamma}_{\alpha} \, \varepsilon^\alpha(x).
\ee
In such spacetimes, therefore, a general diffeomorphism is written in the form
\be
x^\gamma \to x^\gamma + \delta^{\gamma}_{\alpha} \, \varepsilon^\alpha(x).
\ee
Denoting by $Q_T$ the Noether charge relative to $\delta^\gamma$, its integral over a closed spacelike two-dimensional surface $\Sigma$ will be referred to as the Noether charge of $\Sigma$ relative to $\delta^\gamma$. In this case, it can be shown that the entropy $S$ of a Killing horizon is related to Noether's charge according to \cite{WaldEntro,pady2}
\be
S = \frac{2 \pi}{\kappa} \int_\Sigma Q_T
\label{WaldEntro}
\ee
where $\Sigma$ is a two-dimensional surface endowed with a positive-defined metric, allowing in this way the definition of length. It is actually a bifurcate Killing horizon, that is, a surface formed by two Killing horizons that intersect on the space-like surface $\Sigma$. We remark that the subscript `$T$' has been used to remind that the above result holds in a general spacetime that reduces locally to Minkowski, which is transitive under spacetime {\em translations}.

Let us consider now a spacetime that reduces locally to de Sitter, which as we have already seen is transitive under a combination of translations and proper conformal transformations. The Killing vectors corresponding to these transformations are
\be
\xi^\gamma = \xi^{\gamma}_{\alpha} \, \varepsilon^\alpha(x)
\ee
with $\xi^{\gamma}_{\alpha}$ given by Eq.~(\ref{pia}). In this spacetime, therefore, a general diffeomorphism is defined by
\be
x^\gamma \to x^\gamma + \xi^{\gamma}_{\alpha} \, \varepsilon^\alpha(x).
\ee
As a consequence, the Noether charges associated with such transformation will acquire an additional piece related to the proper conformal transformations, as can be seen from Eq.~(\ref{TmK}). This means that in such spacetimes the relation between entropy of a Killing horizon and Noether's charge assumes the form
\be
S = \frac{2 \pi}{\kappa} \int_\Sigma \left[ Q_T - (2l)^{-2} \, Q_K \right]
\label{WaldEntroSL}
\ee
where $Q_T$ represents the part of the Noether charge related to translations, and $Q_K$ the part related to proper conformal transformations. Entropy is consequently made up of two parts, one connected to the translations---which corresponds to the usual, gravitational notion of entropy---and another connected to the proper conformal transformations, which we have called {\em proper conformal entropy}. This is a new concept that is not present in locally-Minkowski spacetimes. In fact, in the contraction limit $l \to \infty$, the underlying de Sitter spacetime reduces to Minkowski and the usual (or Riemannian) expression (\ref{WaldEntro}) for entropy is recovered. 

\section{Black holes in locally-de Sitter spacetimes}

\subsection{General relativity in Cartan geometry}

The replacement of the underlying Poincar\'e-ruled kinematics by a de Sitter-ruled kinematics does not change the dynamics of the gravitational field. There are some differences though. For example, since the cosmological term turns out to be encoded in the local kinematics, it does not appear explicitly in Einstein equation, which on account of the conservation law (\ref{UniCon}), is now written with a new source term\footnote{We use units in which $G = c = \hbar = k_B = 1$.}
\be
R_{\mu \nu} - \onehalf g_{\mu \nu} R = 8 \pi \Pi_{\mu \nu}
\label{EE}
\ee
with $\Pi_{\mu \nu}$ the covariantly conserved current (\ref{TmK}). As an immediate consequence, in contrast to ordinary general relativity, the cosmological term in this theory is no longer required to be constant \cite{evolveL}. Furthermore, the curvature tensor represents both the general relativity {\em dynamic curvature}, whose source is the energy-momentum current, and the {\em kinematic curvature} implied by the underlying kinematics, whose source is the proper conformal current of ordinary matter. In a locally inertial frame---where inertial effects exactly compensate for gravitation---the spacetime metric of any solution of Einstein equation (\ref{EE}) reduces to the de Sitter metric. One can then say that the only difference in relation to ordinary general relativity is the strong equivalence principle, whose new version states that in a locally inertial frame, the laws of physics reduce to that of the de Sitter special relativity. One should note that the {\em local} notion of de Sitter spacetime is different from the usual {\em global}, or homogeneous notion. In fact, considering that the source of the kinematic de Sitter curvature is the proper conformal current of matter, and considering that it is not dynamic, in vacuum it must vanish. The local kinematic curvature can then be thought of as a kind of {\em asymptotically flat} de Sitter spacetime, being in this sense similar to a conjecture made by Mansouri some years ago \cite{mansouri}.

\subsection{Schwarzschild solution}

The Schwarzschild solution in a locally-de Sitter spacetime, although conceptually different, coincides mathematically with the so-called Schwarzschild-de Sitter solution
\be
ds^2 = f(r) \, dt^2 - \frac{dr^2}{f(r)} - r^2 \left(d\theta^2 +
\sin^2\theta \, d\phi^2 \right),
\ee
where
\be
f(r) = 1 - \frac{r^2}{l^2} - \frac{2 M}{r}.
\ee
As can be easily checked, in a locally inertial frame, where gravitation is eliminated by inertial effects, it reduces to the de Sitter metric. Considering furthermore that this metric represents ultimately the Schwarzschild solution, the function $f(r)$ must keep its Schwarzschild form
\be
f(r) = 1 - \frac{2 M}{\mathcal R}
\ee
with
\be
{\mathcal R} = \frac{r}{1 + r^3/2 M l^2}.
\ee
Seen from this perspective, the background de Sitter kinematics is found to produce a change in the Schwarzschild radius, which by equating ${\mathcal R} = 2 M$ turns out to be defined by the solutions of the cubic polynomial equation \cite{3roots}
\be
\frac{r_{SdS}^3}{l^2} - r_{SdS} + 2 M = 0,
\label{CubicFunc}
\ee
where $r_{SdS}$ denotes the radius of the horizons present in the solution.

From now on, for the sake of simplicity, we consider that the de Sitter pseudo-radius is much larger than the Schwarzschild radius: $ l \gg r_{S} = 2M$. One of the three roots of the cubic polynomial equation (\ref{CubicFunc}) is negative, and for this reason it will be neglected. The other two, when expanded in powers of $M/l$, are given by
\be
r_{SdS} = 2 M \Big(1 + \frac{4 M^2}{l^2} + \cdots \Big)
\label{HoriRadius}
\ee
and
\be
r'_{SdS} = l \Big( 1 - \frac{M}{l} - \frac{3 M^2}{2 l^2} + \cdots \Big).
\label{HoriRadius'}
\ee
The first root represents the Schwarzschild horizon, now modified by the underlying de Sitter kinematics. The second solution, on the other hand, represents the de Sitter horizon, which in turn appears modified by the presence of the black hole. {\em This means that the black hole and the de Sitter horizons are connected to each other and cannot be considered separately. In particular, when studying their thermodynamic properties, one has necessarily to consider both horizons as a unique entangled system.} Of course, they have different temperatures, and for this reason they require independent thermodynamic analysis. In the remaining of this section we present such analysis, beginning with the black hole and then considering the de Sitter case.

\subsection{Black hole thermodynamics}

The entropy $S = {A}/{4}$ defined by the de Sitter-modified Schwarzschild radius~(\ref{HoriRadius}) is
\be
S = 4 \pi M^2 \Big(1 + \frac{8 M^2}{l^2} + \cdots \Big).
\label{dSentro}
\ee
The first term on the right-hand side is the usual entropy associated with the translational part of the spacetime local transitivity. The remaining terms, as discussed in Section~\ref{WaldEnSec}, represent the contribution to the entropy coming from the proper conformal part of the spacetime local transitivity.\footnote{One should note that in the exact case only terms proportional to $l^{-2}$ would appear in the entropy expression, as well as in all other thermodynamic variables. In the present case, however, because all variables were expanded in powers of $M/l$, they turn out to be expressed also by an infinite series in powers of $M/l$.} In the formal limit $l \to \infty$, the underlying de Sitter spacetime contracts to Minkowski, and the usual entropy $S = 4 \pi M^2$ of an isolated black hole horizon is recovered. Observe that now all thermodynamic quantities are functions of $M$ and $l$, which represent the two thermodynamic variables of the entangled system \cite{Teitel}. The differential of the entropy (\ref{dSentro}) is consequently given by
\be
dS = 8 \pi M \Big(dM + \frac{16 M^2}{l^2} dM - \frac{8 M^3}{l^3} dl + \cdots \Big).
\label{diffS}
\ee

The horizon temperature, on the other hand, is defined by
\be
T = \frac{\kappa}{2 \pi},
\ee
where $\kappa$ is the surface gravity, which in the case of ordinary Schwarzschild solution has the form
\be
\kappa = \frac{1}{4 M} \equiv \frac{1}{2 \, r_S}.
\ee
In the case of a de Sitter-modified Schwarzschild solution, the surface gravity turns out to be
\be
\kappa \equiv \frac{1}{2 \, r_{SdS}} = \frac{1}{4 M} \Big( 1 - \frac{4 M^2}{l^2} +
\cdots \Big).
\ee
The horizon temperature assumes then the form
\be
T = \frac{1}{8 \pi M} \Big(1 - \frac{4 M^2}{l^2} +
\cdots \Big).
\label{dStempe}
\ee
Similarly to the entropy, the first term on the right-hand side represents the temperature of the usual black hole horizon, whereas the remaining terms represent the change induced by the underlying de Sitter spacetime, which has already discussed is part of the entangled Schwarzschild-de Sitter system.

The de Sitter-modified energy of a black hole can be obtained from the first law of black hole thermodynamics
\be
dE = T \, dS.
\ee
Using expressions (\ref{diffS}) and (\ref{dStempe}), it assumes the form
\be
dE = dM + \frac{12 M^2}{l^2} dM - \frac{8 M^3}{l^3} dl + \cdots.
\ee
An integration yields
\be
E = M + \frac{8 M^3}{l^2} + \cdots \, .
\label{bhE}
\ee
The first term on the right-hand side is the usual energy of a black hole. The remaining terms represent the contribution to the energy coming from the underlying de Sitter spacetime. In the limit $l \to \infty$, the background de Sitter spacetime contracts to Minkowski, and the usual black hole energy is recovered.

\subsection{de Sitter thermodynamics}

Once spacetime reduces locally to de Sitter, there is naturally a de Sitter horizon present in spacetime. We pass now to describe its thermodynamic evolution \cite{G&H}. The entropy $S' = {A'}/{4}$ of the de Sitter horizon with radius $r'_{SdS}$, given by Eq.~(\ref{HoriRadius'}), is
\be
S' = \pi l^2 \Big(1 - \frac{2 M}{l} - \frac{2 M^2}{l^2} + \frac{3 M^3}{l^3} + \cdots \Big).
\label{dSentro'}
\ee
The first term on the right-hand side is the usual entropy associated with the de Sitter horizon. The remaining terms, as discussed in Section~\ref{WaldEnSec}, represent the contribution to the entropy coming from the proper conformal part of the spacetime local transitivity. In the limit $M \to 0$, which represents absence of black hole, these terms vanish and one obtains back the entropy $S = \pi l^2$ of an isolated de Sitter horizon. The differential of $S'$ is
\be
dS' = 2 \pi \Big(l dl - l dM - M dl - {2 M} dM + 
\frac{9 M^2}{2 l} dM + \cdots \Big).
\label{diffSp}
\ee

In the usual case, the temperature of the de Sitter horizon is $T' = 1 / 2 \pi l$, with $l$ the horizon radius. In the presence of a black hole, the de Sitter horizon turns out to be given by $r'_{SdS}$, and the temperature assumes the form
\be
T' \equiv \frac{1}{2 \pi r'_{SdS}} = \frac{1}{2 \pi l} \Big(1 + \frac{M}{l} +
\frac{3 M^2}{2 l^2} + \cdots \Big).
\label{dStempe'}
\ee
The first term on the right-hand side represents the temperature of an isolated de Sitter horizon. The remaining terms represents the change induced by the presence of the black hole.

The black hole-modified energy of a de Sitter horizon can be obtained from the thermodynamic equation
\be
dE' = T' \, dS'.
\ee
Of course, since we interprete the cosmological term as a purely kinematic entity, and not a solution of Einstein's equations with a source possessing negative pressure, no $P' dV'$ appears in the above thermodynamic equation. Using expressions (\ref{diffSp}) and (\ref{dStempe'}), that equation assumes the form
\be
dE' = dl - dM - \frac{3 M}{l} dM + \frac{M^2}{2 l^2} dl + \frac{M^2}{l^2} dM -
\frac{3 M^3}{2 l^3} dl + \cdots \, .
\ee
Integrating we obtain
\be
E' = l - M - \frac{2 M^2}{l} + \frac{13 M^3}{12 l^2} \cdots \, .
\label{dsE}
\ee
The first term on the right-hand side represents the usual energy of an isolated de Sitter horizon. The remaining terms represent the contribution to the energy coming from the presence of a black hole.

\section{Final remarks}

The de Sitter spacetime is usually interpreted as the simplest {\em dynamical} solution of the sourceless Einstein equation in the presence of a cosmological constant, standing on an equal footing with all other gravitational solutions---like for example Schwarzschild and Kerr. However, as a non-gravitational spacetime, the de Sitter solution should instead be interpreted as a fundamental background for the construction of any physical theory, standing on an equal footing with the Minkowski solution. When general relativity is constructed on a de Sitter spacetime, instead of the usual locally-Minkowski Riemannian geometry, spacetime turns out to be endowed with a de Sitter-Cartan geometry \cite{wise,hendrik}. Since Minkowski is transitive under spacetime translations, whereas de Sitter is transitive under a combination of translations and proper conformal transformations, the definitions of diffeomorphism in these spacetimes will change accordingly.

On the other hand, the notion of entropy is directly related to the Noether charges obtained from the invariance of a physical system under spacetime diffeomorphisms. For locally-de Sitter spacetimes, analogously to what happens to all Noether charges---like for example energy---the entropy is found to be composed of two parts: the usual translational-related entropy plus an additional piece related to the proper conformal transformations. The existence of these two kind of entropy was illustrated in the concrete example of the Schwarzschild solution, where the de Sitter length parameter $l$ and the Schwarzschild radius $2M$ are both considered to be thermodynamic variables. In this case, the black hole and the de Sitter horizons are found to make up a unique entangled system.

The temperature of the two horizons are different, which means that the system is not thermodynamically stable \cite{bousso}. This in turn means that there can exist a heat flow from the dynamic black hole horizon to the kinematic de Sitter horizon, and vice versa. In fact, as discussed in Section~\ref{Conser}, in a locally-de Sitter spacetime there is an additional freedom---not present in locally Minkowski spacetimes---that allows {(dynamical)} energy-momentum current to transform into {(kinematical)} proper conformal current, while keeping the {total} energy $E + E'$ constant, where $E$ is the black hole energy (\ref{bhE}) and $E'$ is the de Sitter energy (\ref{dsE}). This conservation law can be written in the equivalent form \[dE + dE' = 0,\] from where the constraint
\be
dl = \left(\frac{3 M}{l} - \frac{13 M^2}{l^2} + \cdots \right) dM
\label{constraint}
\ee
between the thermodynamic variables of the de Sitter and of the black hole horizons can be easily obtained. Such constraint says that every change in the radius of a black hole horizon---either by emitting or absorbing a particle---will produce concomitant changes in the radius of the de Sitter horizon. This result provides a new scenario for the study of cosmology, and in particular for the study of Penrose's conformal cyclic cosmology \cite{PenBook,Penrose}. For example, recent experimental results indicate that the universe expansion became accelerated in the last few billion years \cite{obs1,obs2,obs3}. A possible explanation for this late time acceleration is to suppose that the cosmological term $\Lambda$ is bigger today than it was a few billion years ago. Considering that $\Lambda \sim l^{-2}$, this is equivalent to say that the de Sitter parameter $l$ is becoming smaller, which implies that $dl < 0$. Since the leading-order term of the expansion within parentheses in the constraint (\ref{constraint}) is positive, this implies that $dM < 0$. In this context, therefore, the recent accelerated expansion of the universe can be explained by supposing that in the last few billion years the black holes inside the de Sitter causal horizon---all of them exchanging energy (heat) with the de Sitter horizon---are preponderantly losing more energy (through Hawking radiation) than absorbing (in the form of baryonic matter).

\section*{Acknowledgments}

A.A. would like to thank Universidad Centroccidental Lisandro Alvarado, Barquisimeto, Ve\-ne\-zue\-la, for financial support. J.G.P. thanks FAPESP, CAPES, and CNPq for partial financial support.


\end{document}